# Raman spectroscopy of $(K,Na)NbO_3$ and $((K,Na)_{1-x}Li_x)NbO_3$

H.J. Trodahl[1,2], N. Klein[1], D. Damjanovic[1], N. Setter[1], B. Ludbrook[2], D. Rytz[3] and M. Kuball[4]

[1]*Ceramics Laboratory, Swiss Federal Institute of Technology-EPFL, Lausanne 1015, Switzerland*

[2]*MacDiarmid Institute of Advanced Materials and Nanotechnology, Victoria University, Wellington, New Zealand*

[3]*Forschungsinstitut für mineralische und metallische Werkstoffe Edelsteine/Edelmetalle (FEE), Idar-Oberstein, Germany*

[4]*Applied Spectroscopy Group, H.H. Wills Physics Laboratory, University of Bristol, Bristol BS8 1TL, United Kingdom*

Abstract

A Raman scattering spectroscopy study is reported through all four phases of a $(K_{0.5+\delta}Na_{0.5-\delta})NbO_3$ ($\delta\sim0.03$) crystal and $((K_{0.5+\delta}Na_{0.5-\delta})_{1-x}Li_x)NbO_3$ for x=0.02 and 0.03. The transitions among the ferroelectric phases in the Li-free crystal are homogeneous and strongly hysteretic, with no evidence of a mixed-phase region between the rhombohedral and orthorhombic phases. The Raman pattern in the lowest-temperature phase in the Li-doped material differs significantly from that found for x=0 and suggest a structural phase boundary exists at for a Li concentration of less than 2%.

Potassium/sodium niobate, $K_{1-x}Na_xNbO_3$ (KNN), has been investigated for its ferroelectric properties for many decades,[1-5] but has enjoyed a renaissance in the past few years as an environmentally-friendly potential alternative for $Pb(Zr_xTi_{1-x})O_3$ (PZT) in the ubiquitous applications of its strong piezoelectric response in actuators, sensors and ultrasonic transducers. Substitution of Li on the A site and Ta on the B site, $(K_{0.5}Na_{0.5})_{1-x}Li_xNb_{1-y}Ta_yO_3$ (KNLN for y=0), suggested a morphotropic phase boundary (MPB) with strong piezoelectric properties might mimic the behavior of PZT.[6-10] In contrast to PZT, however, this boundary is now recognized to be strongly temperature dependent.[11,12] The compositions with the highest piezoelectric properties exhibit the polymorphic phase transition between tetragonal and orthorhombic phases in KNLN close to the room temperature.[11,13] Other lead-free materials (e.g. $Na_{0.5}Bi_{0.5}TiO_3$) considered as alternatives to PZT also exhibit complex composition/temperature phase diagrams.[13] It is central in this search for a lead-free piezoelectrics to identify the temperature-dependent phase diagrams for the promising materials. Several recent papers have addressed this question in KNLN and related materials through the use of Raman spectroscopy.[11,12,14-18] In particular we recently noted anomalies in the Raman spectra of KNLN that raised questions about whether the low-temperature phase transitions in KNLN are in accordance with those in KNN.[12] However, there is no report in the literature of the Raman spectra across the full set of four phases for KNN. In this paper we report full temperature-dependent Raman spectra for single-crystalline KNN and for KNLN. A comparison of the transition temperatures and spectra reveals a surprising difference in the low temperature phases, implying an altered low-temperature crystal structure for as little as 2% Li substitution in KNN.

Extensive early studies [1-5] established that, for $0 < x < 0.55$, $K_{1-x}Na_xNbO_3$ undergoes the same series of phases (cubic-tetragonal-orthorhombic-rhombohedral) (C-T-O-R) with falling temperature that is common to many perovskite ferroelectrics, and in particular for $KNbO_3$ (KN). The phase diagram develops smoothly from x = 0 to 0.55, and there can be little doubt that the structural changes associated with these phases are entirely analogous to those in KN.

The KNN crystal used in the present study were grown by the top seeded solution growth (TSSG), at 1093°C, with cooling rate of 0.15 °C/h. The seed orientation was <001> and the crystal has a milky appearance with a high domain density. The approximate K:Na ratio measured by inductive plasma spectroscopy-mass spectrometry was 0.59:0.41 while the EDX study gave 0.57:043. Both compositions are on the K rich side of the morphotropic phase boundary in the $KNbO_3$-$NaNbO_3$ phase diagram at about K:Na ratio of 55:45 (x=0.45).[2] A KNLN (3%) crystal was also grown by the TSSG, with the K:Na ratio of 45:55 (x=0.55), very close to the reported phase boundary to a low-temperature monoclinic phase in Li-free material.[1-5] In order to separate the effects of varying K:Na and Li concentration we performed Raman studies of two ceramic samples with K:Na of 50:50 and Li concentrations of 2 and 3%. The preparation and treatment of these samples has been described previously.[12]

For Raman studies the samples were placed in a liquid nitrogen-cooled Linkam THMS600 cold/hot stage. Unpolarised back-scattered Raman spectra were collected with either a Renishaw InVia or a Jobin Yvon LabRam HR spectrometer using the 488 and 514 nm Ar+ lines. Notch filters at the entrance to the spectrometers prevented access to the soft mode below 100 cm$^{-1}$. The incident light was focused to a spot of about 500 nm diameter and the power was held below 1 mW, which did not significantly raise the sample temperature. Studies to lower temperatures, carried out using a liquid He flow-through cryostat, confirmed that there were no phase transitions below 80 K.

Figure 1 shows typical R- and O-phase spectra at 125 K, O- and T-phase spectra at 465 K, and T- and C-phase spectra at 690 K of KNN. (The hysteretic nature of these first-order transitions permits displaying spectra in adjacent phases at a common temperature.) The spectral shifts seen in Fig. 1 clearly signal the phase changes, with the detailed phonon frequencies from multiple-line fitting plotted as a function of temperature in Figure 2. The discontinuous shifts at the transition temperatures are clear the R-O and O-T boundaries. Across the T-C boundary the changes are primarily in the widths of the lines, as is seen clearly in Fig. 1a. The trends within each phase consist primarily of mode softening with rising temperature, exactly as expected to follow from thermal expansion. Note, however, one exception; the line near 425 cm$^{-1}$ anomalously hardens with temperature throughout the O phase.

Table I lists the transition temperatures measured for both increasing and decreasing temperatures, showing excellent agreement with transition-temperature data from the literature.[2] All of the transitions displayed hysteresis, with the largest (>50%) at the R-O transition and the smallest (~5 K) at the T-C transition. Within the hysteretic temperature range the spectra were typical of the higher-temperature phase while lowering the temperature and of the lower-temperature phase while raising the temperature. There were no mixed-phase spectra at any temperature within that range, clearly signaling homogeneous transitions at the extremes of the range.

|       | (a) rising | (b) falling | (c) ref. 2 |
|-------|------------|-------------|------------|
| R/O   | 158 K      | 98 K        | 127 K      |
| O/T   | 470 K      | 455 K       | 473 K      |
| T/C   | 690 K      | 685 K       | 680 K      |

Table I. Transition temperatures through the ROTC phase sequence in KNN. Columns (a) and (b) are the transition temperatures determined while the temperature is rising and falling, respectively, compared in column (c) with historical data based on dielectric measurements.

The close similarity of the KNN Raman spectra of Fig. 1 and those in KN permit a confident assignment of the features at temperatures below 500 K, based on Raman and infrared studies and vibrational calculations for the zone-centre phonons in KN across all of its phases.[18-22] These zone-centre assignments differ from zone-boundary internal modes proposed for some lines in KNN,[17] which could be rendered Raman active only by strong disorder. Indeed these materials are subject to disorder-induced scattering at higher temperature as will be discussed in the next paragraph, but the narrow lines in the R and O phases and the lower temperatures of the T phase are zone-centre modes, as they are in KN.

The strong broad Raman bands peaking at 250 and 600 cm$^{-1}$ in the cubic phase, as shown in the C-phase data in Fig. 1a, are a clear signal of disorder at these elevated temperatures. Raman activity in well-ordered cubic perovskites is symmetry forbidden. Nonetheless these features are common in the cubic phase of pervoskite ferroelectrics and they signal a structure with local static and/or dynamic disorder reminiscent of the distortions in ferroelectric phases.[23] A disorder-induced background is found also in the spectra of the higher-temperature section of the tetragonal phase, but the discontinuous loss of intensity and increased line width of the Raman bands across the tetragonal-cubic transition seen in Fig. 2a signals a clear coherence-length collapse of the tetragonal distortion.

A recent XRD study of ceramic KNN suggested that the O-R transition is inhomogeneous, initiated by nano-islands of rhombohedral material within an orthorhombic matrix, and that the transition is completed only after the islands grow to coalescence.[17] There is no evidence of any such gradual phase transition in the present data; in contrast the transition is immediate, discontinuous and complete, and as discussed above it shows a very large hysteresis. It is unclear whether the apparent disagreement between the studies is fundamental or merely related to the relatively small crystallites in the XRD study.

Turning next to KNLN (2%) ceramic we find phase transitions near 220 K and 450 K, as we reported recently.[12] It is tempting to simply accept these as the R-O and O-T transitions as in KNN. Note, however, that the lowest temperature transition occurs at a substantially higher temperature than the R-O transition in KNN. Furthermore the Li-concentration dependence of that transition temperature in KNLN ceramics seen in Ref 12 is very weak, and it extrapolates to 250±20 K at zero Li concentration, more than 100 K higher than the R-O transition in KNN. There is thus a question as to whether the ground state in KNLN is a rhombohedral phase analogous to that in KNN. In Figure 3 we compare the "R"-, O- and T-phase spectra in the two samples, seeking similarity that might support a rhombohedral assignment for the lowest-temperature phase of KNLN. The spectral differences seen in the O and T phases are no more than shifts in the frequencies of the features, while in the lowest-temperature phase there are both larger frequency shifts and a loss of a satellite on the low-frequency side of the 280 cm$^{-1}$ line in the low-temperature phase of KNLN. An even more striking difference is found in the shift of the 280 cm$^{-1}$ line in KNLN, which softens across the transition from the low-temperature to orthorhombic phases as seen clearly in Figure 4. That softening is in disagreement with the general stiffening with increasing temperature across all transitions in KNN and shown in Fig. 1. It is clear that the crystal structure of KNLN is very different from KNN, despite only 2% A-site Li substitution.

It is significant that the Raman spectra from our KNLN (3%) ceramic and the Na-rich KNLN (3%) crystal are all but indistinguishable from the data shown in Figs 3 and 4. We draw the inference that the structures if these solid solutions are at most a weak function of the K:Na ratio across the 45:55-55:45 range, but that there is a phase transition in the ground state that occurs between 0 and 2% Li concentration.

In summary we have reported a Raman spectroscopic study through all of the phases of a $(K_{0.5+\delta}Na_{0.5-\delta})NbO_3$ crystal for $\delta \sim 0.05$. The spectra are similar to those reported in $KNbO_3$, with only modest shifts in the frequencies of the prominent lines, and confirm that KNN and KN adopt the same series of crystal structures through all of their ferroelectric phases. There is clear hysteresis at each transition, with no evidence of a mixed-phase region between the rhombohedral and orthorhombic phases of KNN. A similar close correspondence to KN is found in KNLN spectra for the tetragonal and orthorhombic phases. In contrast, we find that even a minimal 2% Li substitution leads to a lowest-temperature phase with Raman spectra that differ from the rhombohedral-phase spectra of KNN and KN. It is clear that the entire region near K:Na of 50:50 and for Li concentrations below 2% require further exploration to investigate the phase boundary these results suggest occurs between 0 and 2% Li.

Acknowledgements

Work was in part carried out in the framework of the European FP6 CRAFT project, IMMEDIATE. The MacDiarmid Institute is supported by the New Zealand Centre of Research Excellence programme. The authors are grateful to CIME-EPFL for the SEM investigations, and for access to the mass spectrometer facilities at the University of Lausanne.

Figure captions

Figure 1. (Color online) Discontinuous spectral changes across three phase transitions in KNN. The spectra are compared at a common temperature in neighboring phases, a comparison permitted by hysteresis.

Figure 2. (Color online) Temperature dependence of the major Raman lines in KNN across all four phases. Data in the R phase are shown as inverted red triangles, in O as black circles, in T as blue squares and in C as upright magenta triangles. As indicated by the arrows, pure R-phase spectra were recorded when raising the temperature from 90 to 155 K, and pure O-phase spectra while lowering from 160 to 95 K. Similar single-phase spectra were observed in the hysteretic range between the O and T phases.

Figure 3. (Color online) Comparison of the spectra in the three ferroelectric phases of KNN and KNLN.

Figure 4. (Color online) Spectral changes across the lowest temperature phases in KNLN. These results can be compared to the equivalent data for KNN shown in Fig. 1c.

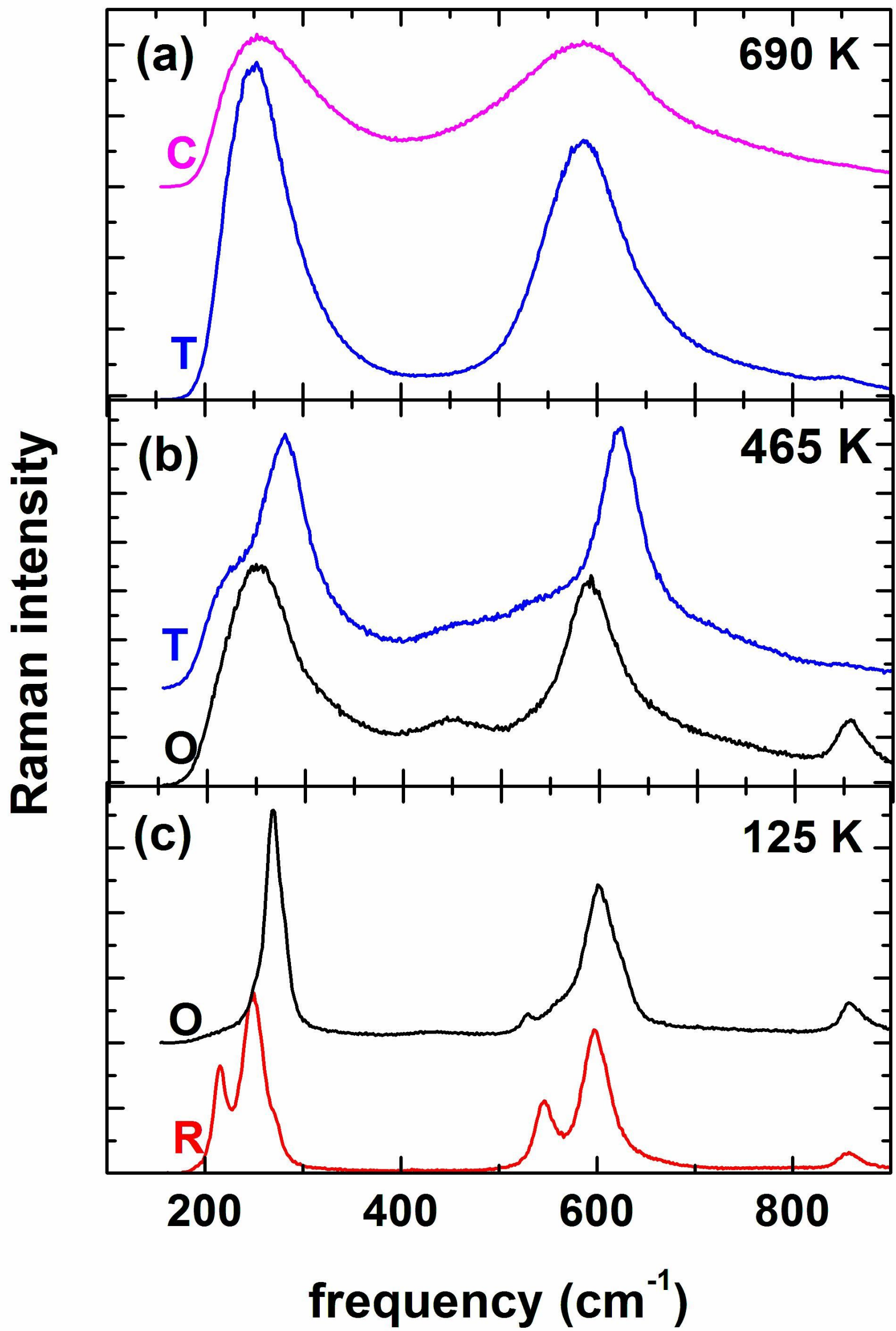

(a) 690 K

C

T

(b) 465 K

T

O

(c) 125 K

O

R

Raman intensity

frequency (cm⁻¹)

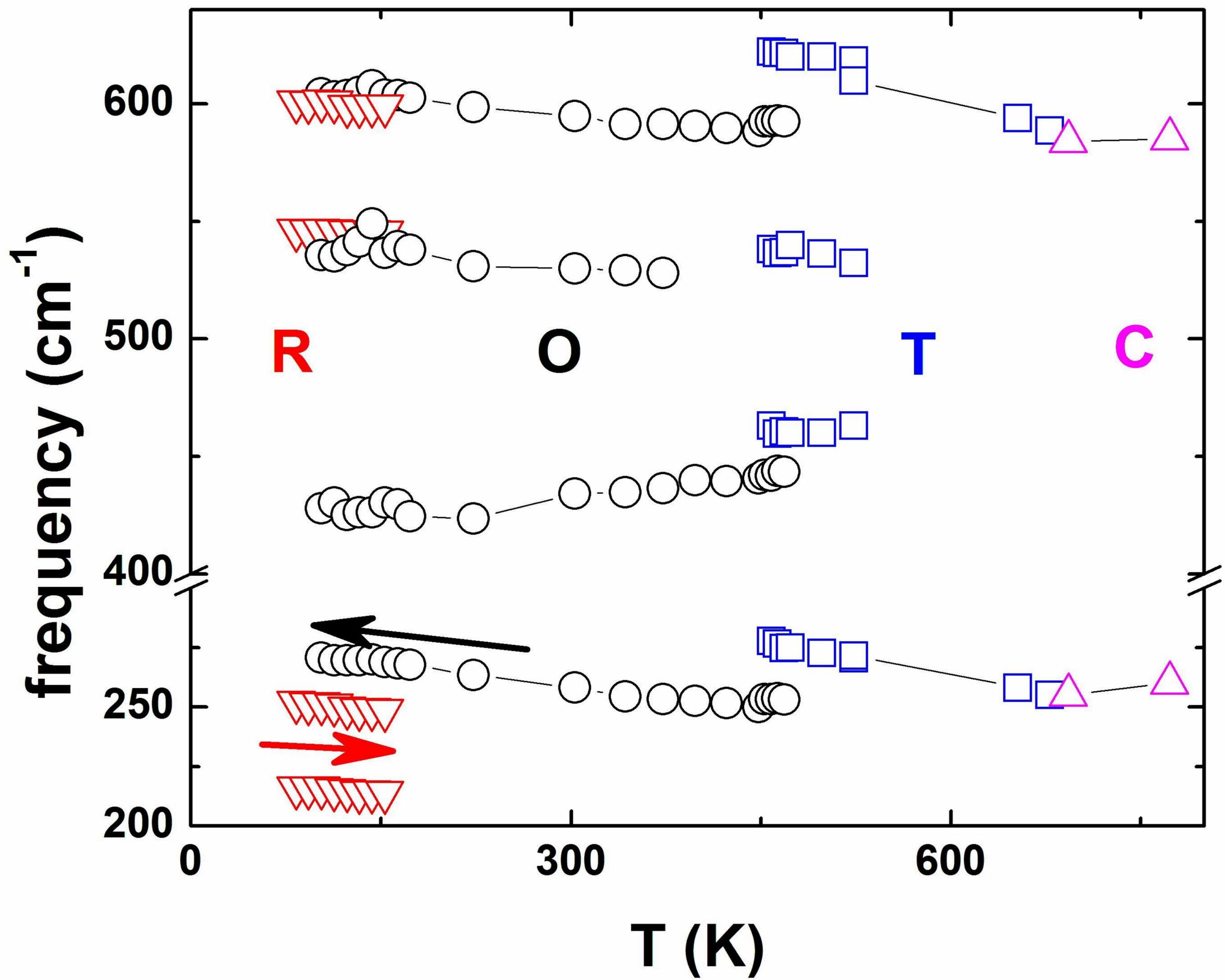

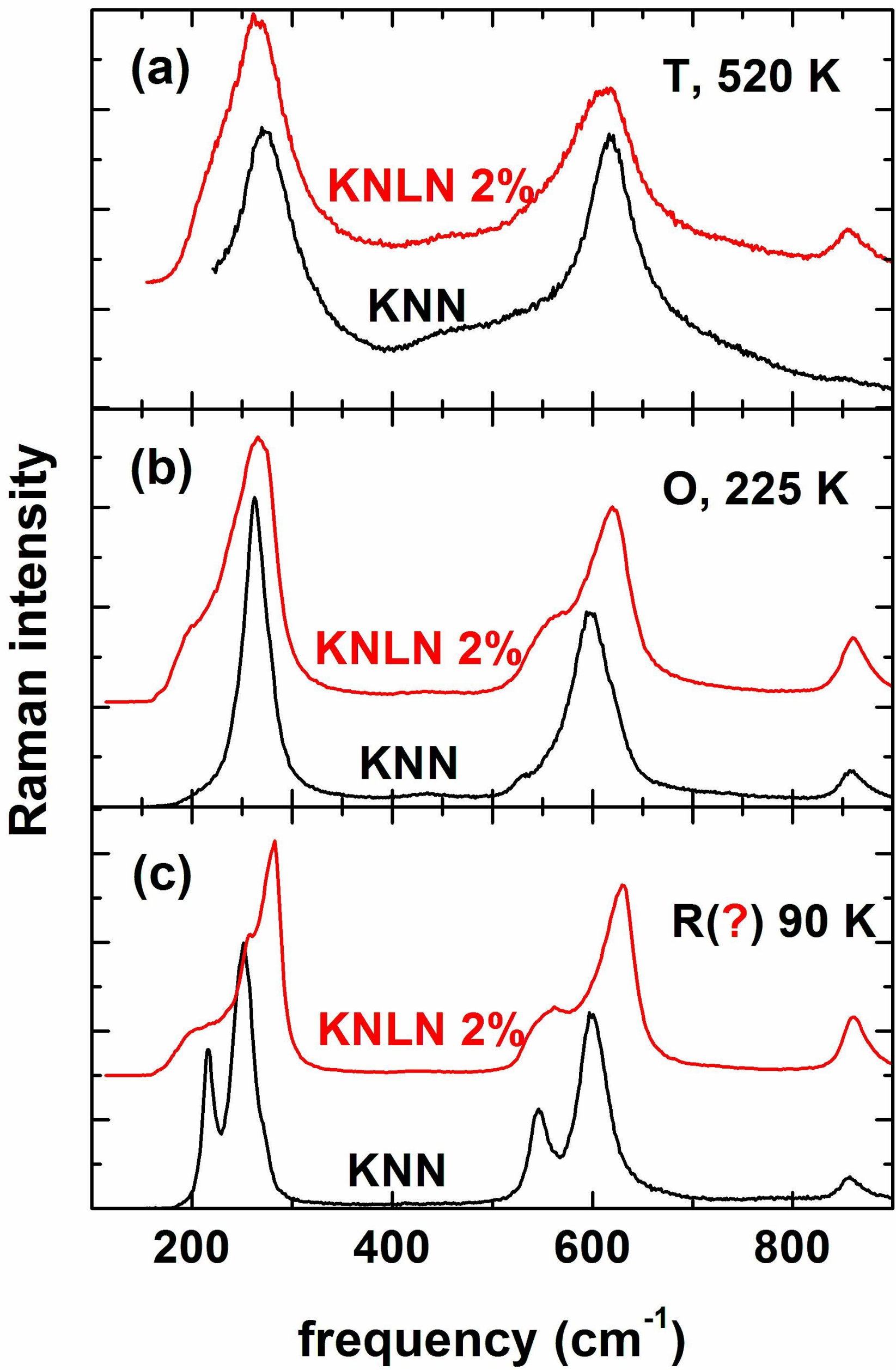

(a) T, 520 K
KNLN 2%
KNN

(b) O, 225 K
KNLN 2%
KNN

(c) R(?) 90 K
KNLN 2%
KNN

Raman intensity

frequency (cm$^{-1}$)

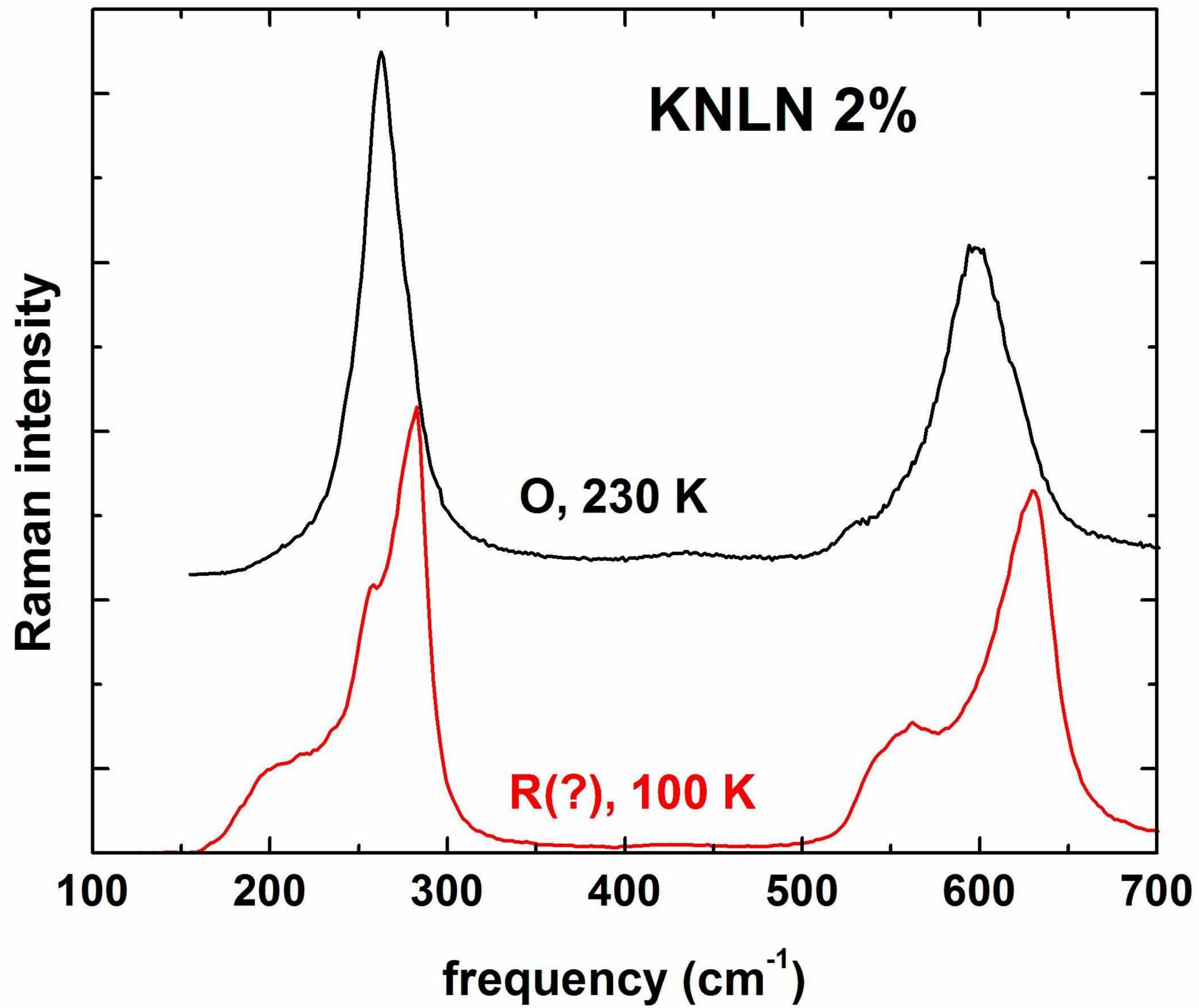